\documentclass[10pt,showpacs,twocolumn]{revtex4}
\usepackage{color}
\usepackage{graphicx}
\usepackage{amsmath}
\usepackage{amssymb}
\usepackage{}
\begin{document}

\preprint{APS/123-QED}

\title{High-order harmonic generation of doped semiconductor}

\author{Tengfei Huang$^{1}$, Xiaosong Zhu$^{1,}$\footnote{zhuxiaosong@hust.edu.cn}, Liang Li$^{1}$, Xi Liu$^{1}$, Pengfei Lan,$^{1,}$\footnote{pengfeilan@mail.hust.edu.cn} and Peixiang Lu$^{1,2,}$\footnote{lupeixiang@mail.hust.edu.cn}
}

\affiliation{%
 $^1$School of Physics and Wuhan National Laboratory for Optoelectronics, Huazhong University of Science and Technology, Wuhan 430074, China\\
 $^2$Laboratory of Optical Information Technology, Wuhan Institute of Technology, Wuhan
430205, China\\
}%

\date{\today}

\begin{abstract}
We investigate the high-order harmonic generation (HHG) in doped semiconductors. The HHG is simulated with the single-electron time-dependent Schr\"{o}dinger equation (TDSE). The results show that the high-order harmonics in the second plateau generated from the doped semiconductors is about 1 to 3 orders of magnitude higher than those from the undoped semiconductor. The results are explained based on the analysis of the energy band structure and the time-dependent population imaging. Our work indicates that doping can effectively control the HHG in semiconductor.
\end{abstract}

\pacs{42.65.Ky, 42.65.Re, 32.80.Rm, 42.50.Hz}

\maketitle

\section{Introduction}
With the fast development of laser technology, the interaction between intense laser pulses and matters has been studied extensively over the past several decades and revealed many interesting phenomena \cite{Krausz,Corkum,ZWang}. One of the most interesting phenomena is high-order harmonic generation (HHG) \cite{Krausz2,Schafer}, because it promises many important and unprecedented applications like generating coherent ultrafast extreme violet radiations \cite{Krausz,Paul,Hentschel,FWang} and probing the ultrafast dynamics of atomic, molecular and solid systems \cite{Uiberacker,Schiffrin,Leone,Lan,DWang}.

Recently, high-order harmonics emitted from the bulk solids were detected \cite{Ghimire}. Compared with the gas, the solid has periodic structure and high density  \cite{Ghimire2,Luu}. Therefore, it has potential to produce more efficient HHG than the gas \cite{Vampa}. By analyzing the spectrum of the solid HHG, it is possible to study the structures of solid materials \cite{Vampa,Yu,Jiang,Kemper}. The solid HHG can also provide a new path to investigate the attosecond electron dynamics in solid materials \cite{Schultze} and to reconstruct the energy-band structures of solid crystals \cite{Vampa2}.

Recent works have shown that the solid HHG has two-plateau structure \cite{Ndabashimiye,Wu,Wu2,Du}. The intensity of the second plateau is about five orders of magnitude lower than that of the primary plateau \cite{Wu}. One-band model \cite{Pronin} and multi-bands model \cite{Hawkins,Hawkins2,Vampa3,McDonald} are used to explain the mechanism behind HHG. It is considered that the two-plateau structure is attributed to the multi-bands structure of the solid \cite{Hawkins,Hawkins2}. Wu $et\ al.$ \cite{Wu2} suggest that the interband current between the valence band and the first conducting band contributes the primary plateau of the high-order harmonic spectrum, and the interband current between the valence band and the second to the third conducting bands contributes the second plateau. On the other hand, the electron-hole recollision model in wave vector k space is proposed to explain the mechanism of HHG in soilds \cite{Du,Vampa,Vampa3,McDonald}. It is suggested that the first plateau arises from electron-hole recollision, while the higher plateaus arise from dynamic Bloch oscillations.

Nowadays, in energy band engineering, doping is widely used to improve the physical, e.g electrical, magnetic and optical characteristics of objects through changing the energy band structures of the target \cite{Bryan}. It indicates that the characteristics of HHG in solids can be controlled by doping \cite{Cox}. In this paper, we investigate HHG in the doped semiconductors and discuss the influence of doping on HHG. The results show that the energy bands are changed by doping and the intensity of the second plateau of HHG is improved by about 1 to 3 orders of magnitude compared with that of the undoped semiconductor. This result is analyzed based on the picture of energy bands and the time-dependent population imaging.

\section{Theoretical model}
In our work, we investigate the laser-crystal interaction and HHG in the doped and undoped semiconductor by sovling the time-dependent Shr\"{o}dinger equation (TDSE). The laser field is polarized along the $\vec{x}$ axis. In the length gauge, the time-dependent Hamiltonian is written as

\begin{eqnarray}
\hat{H} = \hat{H}_0+xE\left ( t \right )
\end{eqnarray}

\noindent where $\hat{H}_0=\hat{p}^{2}/2+v(x)$. Atomic units are used in this paper unless otherwise stated. $v(x)$ is the periodic potential of the lattice. In this work, the undoped semiconductor is modeled by Mathieu-type potential \cite{Slater} $v(x)=-v_0[1+\cos(2\pi x/a_0)]$, with $v_0=0.37$ a.u. and the lattice constant $a_0=8$ a.u. The Mathieu type potential has been widely used to simulate the optical lattice \cite{Breid,Chang} and solid HHG \cite{Wu2,Guan,Du2}.

For the doped semiconductor, we discuss the situation that the dopant replaces the atom of the undoped semiconductor peoriocally. We assume that the dopant will not change the lattice constant. The potential based on Mathieu-type potential is written as

\begin{equation}
v=\left\{
\begin{array}{rcl}
-v_{0}[1+\cos (2\pi x/a_{0})] && {a\leq x\leq b \quad or \quad c \leq x \leq d},\\
-v_{1}[1+\cos (2\pi x/a_{0})] && {b < x < c}.
\end{array} \right.
\end{equation}

\noindent The potentials of the doped (the blue dashed line) and undoped (the red solid line) semiconductor in a repetitive unit are shown in Fig.1. The potential parameter of the dopant between $b$ and $c$ is $v_1=0.52$ a.u. One can see five atoms in the region $[a,d]$. Between $b$ and $c$, the potential parameter of the origin atom $v_{0}$ is replaced by that of the dopant atom $v_{1}$. So the doping rate is 0.2.

\begin{figure}[htb] 
	\centerline{
		\includegraphics[width=9cm]{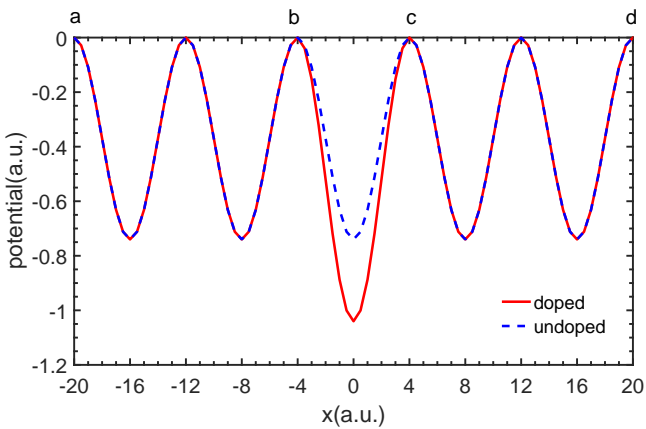}}
	\caption{The blue dashed line shows the potential of 1-D undoped semiconductor. The red solid line shows the potential of 1-D doped semiconductor.}
\end{figure}

To obtain the energy bands of the doped and undoped semiconductors, we solve the eigenvalue equation of $\hat{H}_0$

\begin{eqnarray}
\hat{H}_0\varphi_n\left( x\right) = E_n\varphi_n\left( x\right),
\end{eqnarray}

\noindent where $n$ is the eigenstate number and $\varphi_n\left( x\right)$ is the eigenstate wave function. To solve the eigenvalue equation, we diagonalize $\hat{H}_0$ on a coordinate grid \cite{Liu}. With the finite-difference method, the operator $\hat{H}_0$ is represented by an $N\times N$ matrix {\bf H}, where $N$ is the number of the grid points. The nonzero elements of the matrix {\bf H} are given by

\begin{equation}\label{eq:12}
\begin{split}
{\bf H}_{i,i} & = \frac{1}{(dx)^{2}}+V_{i},  \\
{\bf H}_{i,i+1} & = - \frac{1}{2(dx)^{2}}, \\
{\bf H}_{i+1,i} & = {\bf H}_{i,i+1}, \\
\end{split}
\end{equation}

\noindent where the grid spacing $dx$ is 0.25 a.u. and $V_{i}$ is the ith element of the one-dimensional grid of $v(x)$. The results of both doped and undoped semiconductors are calculated in the real space within the region $[0, 4000]$ a.u. (500 lattice periods).

Figure 2 shows the energy band structure of the doped and undoped semiconductors. The red circles and the black points correspond to the doped semiconductor and the undoped semiconductor respectively. Each energy band of the undoped semiconductor can be distinguished clearly. The valance band (VB) and conducting bands (CB1, CB2 and CB3) correspond to the state number 501-1000, 1001-1500, 1501-2000 and 2001-2500. The energy gap between VB and CB1 is labeled as e1 and the energy gap between CB1 and CB2 is labeled as e2. It is shown in Fig.2(b) that the band gaps e1 and e2 of the doped semiconductor are narrower than those of the undoped semiconductor. 

\begin{figure}[htb]
	\centerline{
		\includegraphics[width=8.8cm]{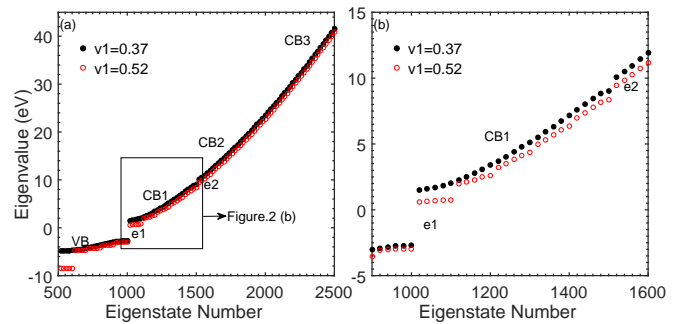}}
	\caption{(a) shows the energy band structures of the doped and undoped semiconductors. The black points show the eigenvalue of the undoped semiconductor, while the red circles show the eigenvalue of the doped semiconductor. We choose 1 point every 20 points.(b) shows the band structure region that includes the gaps e1 and e2.}
\end{figure}

Doping changes the periodicity of the semiconductor. Therefore, compared with the undoped semiconductor, the energy bands of the doped semiconductor are splited into small bands and the small energy bands corresponding to state number 1-100, 501-600 and 1001-1100 locate away from other energy bands. As a result, the energy gaps e1 and e2 become much narrower in the doped semiconductor.

To obtain the time-dependent wave function $\psi\left( t\right)$, we solve the TDSE using the second-order split-operator method

\begin{equation}\label{eq:12}
\begin{split}
\psi\left( t+dt\right)  & = exp\left( - \frac{idt\hat{T}}{2}\right)  \\
 & \times exp\left(  -idt\hat{V}\right)  \\
 & \times exp\left( -\frac{ idt\hat{T}}{2}\right)\psi\left( t\right)+O(dt^3) \\
\end{split}
\end{equation}

\noindent where $\hat{T}=\hat{p}^{2}/2$ and $\hat{V}=v(x)+xE(t)$. The number of the time points is $10000$. In Eq.(5), commutation errors give rise to the third order term in $dt$. The Eq.(5) is solved by the spectral method \cite{Feit}. We solve the first and the third term on the right hand side in momentum space by the fast Fourier transform algorithm. The second term are multiplied directly in position space. In this work, we adopt a sine-squared envelope for the driving laser pulses with the total duration of eight optical cycles ($t_{c}$). The wavelength of the driven laser is $3.2$ $\rm \mu m$ and the intensity is $8.087\times10^{11}$ $\rm W/cm^2$. We choose the eigenstate populated at the top of the valence band as the initial state \cite{Wu2,Guan}. To overcome the unphysical reflections of the wave function $\psi\left( t\right)$ at the edges of the grid spacing, we use a $cos^{\frac{1}{8}}$ absorbing boundary. The width of the absorbing boundary is $kL$, where $k=0.0833$ is the scale and the $L=4000$ a.u. is the length of the real space we used. We find that the $k$ is small enough to insure the results are stabilized Since the lowest band is very flat and deeply bound, it plays a negligible role in the HHG dynamics. 

With the time-dependent wave function $\psi\left( t\right) $, the laser-induced current can be obtained as

\begin{eqnarray}
j\left( t\right) = -\langle\psi\left( t\right) |\hat{p}|\psi\left( t\right) \rangle.
\end{eqnarray}

\noindent We multiply $j(t)$ by a Hanning window \cite{Wu2}. The harmonic spectrum is obtained by calculating the Fourier transform of the laser-induced current

\begin{eqnarray}
H\left( \omega\right) \propto | j\left( t\right) e^{i\omega t} dt |^{2}.
\end{eqnarray}

\noindent To better analyze the HHG in the doped and undoped semiconductors, the time-dependent population imaging (TDPI) is also calculated \cite{Liu}. To obtain the TDPI, the instantaneous population $C_{n}(t)$ on each eigenstate is calculated by the modulus square of the time-dependent projection of $\psi\left( t\right) $ on $\varphi_n$ as

\begin{eqnarray}
|C_{n}\left( t\right) |^{2} = \langle\varphi_{n} |\psi_{n}\left( t\right) \rangle^{2}.
\end{eqnarray}

\noindent The corresponding eigenvalue of $\varphi_{n}$ is $E_{n}$. $|C_{n}\left( t\right)|^{2} $ can be understood as the time-dependent probability of electrons occupying on the eigenenergy $E_{n}$. Then the TDPI picture is obtained by plotting $|C_{n}\left( t\right)|^{2} $ as a function of time $t$ and eigenenergy $E_{n}$.

\section{Results and Discuss}

Figure 3 shows the HHG spectrum of the doped (blue solid line) and undoped (red dash-dotted line) semiconductors. The HHG spectrum of the undoped semiconductor shows a two-plateau structure. The first plateau starts at the 18th order and has a cut-off at the 32nd order. The second plateau starts at the 40th order and has a cut-off at the 112nd order. For the HHG spectrum of the doped semiconductor, the intensities of the high-order harmonics between the 10th and the 32nd orders are lower than those of the undoped semiconductor, while the intensities of the high-order harmonics between the 40th and the 80th orders are about 2 orders of magnitude higher than those of the undoped semiconductor. The second plateau is dramatically enhanced by the doping.
\begin{figure}[htb]
	\centerline{
		\includegraphics[width=9.5cm]{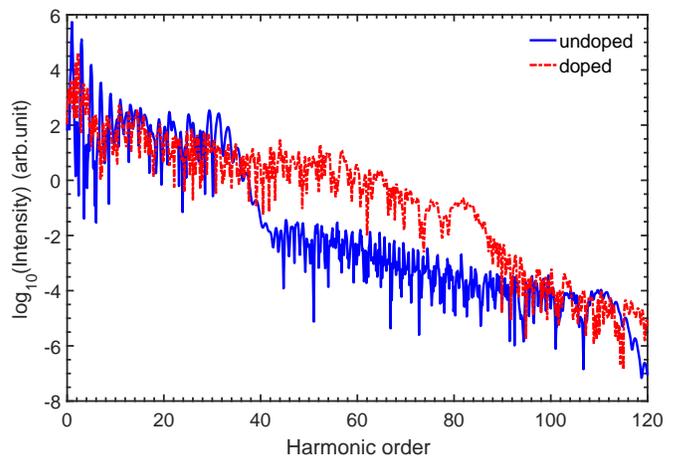}}
	\caption{The blue line shows the HHG spectrum of the undoped semiconductor. The red dash-dotted line shows the HHG spectrum of the doped semiconductor.}
\end{figure}

\begin{figure*}[htb]
	\centerline{
		\includegraphics[width=12cm]{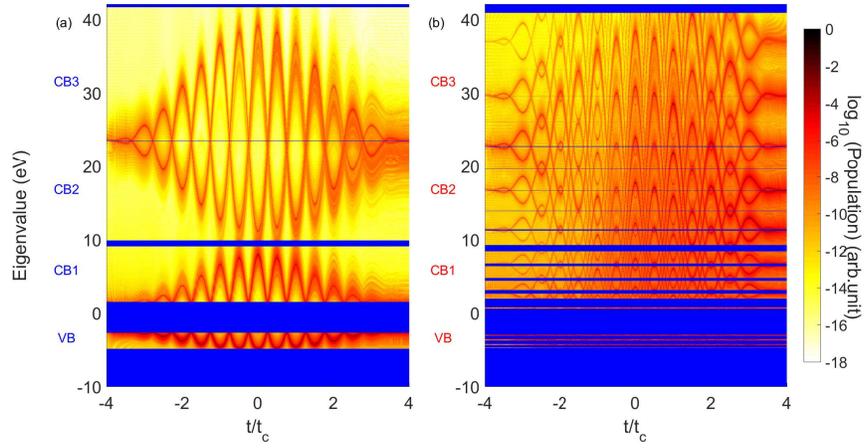}}
	\caption{(a) shows the TDPI picture of the undoped semiconductor model. (b) shows the TDPI picture of the doped semiconductor model. Blue areas are band gaps.}
\end{figure*}

\begin{figure}[htb]
	\centerline{
		\includegraphics[width=8cm]{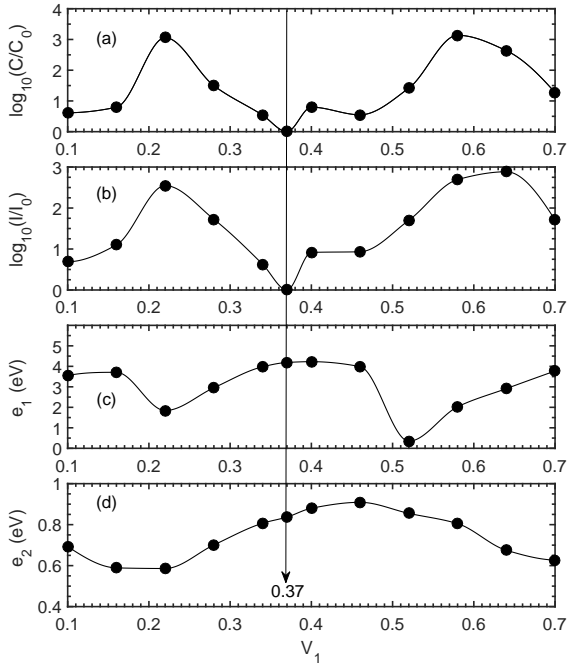}}
	\caption{(a) shows the comparison between the total population of CB2 and CB3 in the doped semiconductors with that of the undoped semiconductors. (b) shows the comparison of average intensity of second plateau (between 40th and 80th orders) between the doped and undoped semiconductors. (c) shows the band gap e1 between VB and CB1 of different doped semiconductors. (d) shows the band gap e2 between CB2 and CB3 of different doped semiconductors.}
\end{figure}

To understand the mechanism of the enhancement of the second plateau intuitively, we show the TDPI picture in Fig.4. Figure 4(a) shows the electron populations in the energy bands of the undoped semiconductor. They are driven forth and back in each energy band by the external laser field. The oscillations of electron populations in the energy bands correspond to the laser-driving Bloch oscillations of electrons in reciprocal space. Figure 4(b) shows the electron populations in the energy bands of the doped semiconductor. Besides the oscillations of electron populations, one can also see the reflection of electron populations at the border of the energy bands, i.e. at the border of the Brillouin zone in Fig.4(b). This is because the energy bands of the doped semiconductor are separated into small bands. As the doping rate is 0.2, the Brillouin zone of the doped semiconductor become 0.2 as that of the undoped semiconductor. Therefore it is easier for electrons to oscillate to the border of the Brillouin zone of the doped semiconductor. And at the border of the Brillouin zone, the electrons can be reflected in the same band or be much more easily to tunnel to higher bands \cite{Li}. So the electron populations in the CB2 and CB3 of the doped semiconductor are higher than those of the undoped semiconductor.

The phenomenon mentioned above can also be explained by the band structure of the doped semiconductor. When the laser field interacts with the semiconductor, the electrons in the VB begin to oscillate in the same band and have possibility to tunnel to energy band CB1 through the band gap e1. We label it as the process 1. After they populate on the CB1, there are two possible paths. On the one hand, some of them can oscillate to the border of the Brillouin zone and tunnel to higher energy bands CB2 and CB3 through a narrow band gap e2. Then these electrons in CB2 and CB3 can transfer to VB and radiate high-order harmonics in the second plateau. We label it as the process 2. On the other hand, some of the electrons in CB1 can transfer back to VB and radiate high-order harmonics in the first plateau. In Fig.4(a), only a small portion of electrons of the undoped semiconductor can arrive at the top of CB1 (i.e. the border of the Brillouin zone), so the process 1 and the process 2 are weak. Thus the electron populations in the CB2 and CB3 are small and the intensity of the second plateau is low. On the contrary, in Fig.4(b), because the energy bands become smaller, most of the electrons can populate on the top of the small energy bands and tunnel to higher energy bands. Besides, it is shown in Fig.2(b) that doping makes the band gaps e1 and e2 of the doped semiconductor narrower than those of the undoped semiconductor. Therefore, the process 1 and the process 2 are strengthened. Consequently, the electron populations in the CB2 and CB3 of the doped semiconductor are larger. And the second plateau of the doped semiconductor is about 2 orders of magnitude higher than that of the undoped semiconductor. On the other hand, the smaller Brillouin zone and the narrower band gap e1 can strength the populations in CB1. While the narrower band gap e2 can make more populations in CB1 transfer to CB2, resulting in less transitions from CB1 to VB. So the intensity of the first plateau is about 1 order of magnitude lower than that of the undoped semiconductor. 

\begin{figure}[htb]
	\centerline{
		\includegraphics[width=8cm]{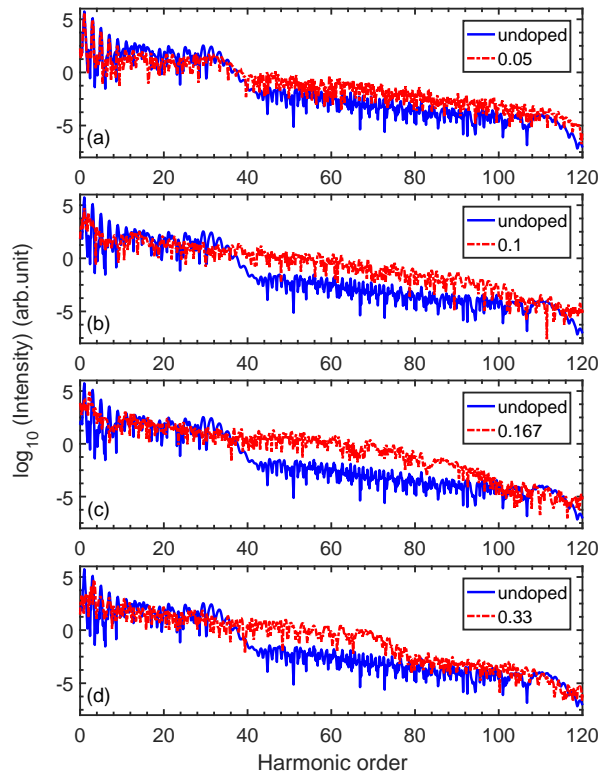}}
	\caption{The blue lines in each figure show the HHG spectrum of the undoped semiconductor. The red dash-detted lines show the HHG spectrum of the doped smiconductors under different doping rates 0.05, 0.1, 0.167 and 0.33 in Fig.6(a), (b), (c), (d). $v_{1}=0.52$ a.u.}
\end{figure}

To further investigate the effect of doping, we compare the populations on CB2 to CB3 of the doped semiconductors with different values of $v_{1}$. The results are shown in Fig.5. In Fig.5(a), $C$ is the total electron population of the energy bands CB2 to CB3 in the doped semiconductor, and $C_{0}$ is the population of the energy bands CB2 and CB3 in the undoped semiconductor. Figure 5(b) shows the comparison between the average intensities in the second plateau (between 40th and 80th orders) of the doped and undoped semiconductors. $I$ is the average intensities of the high-order harmonics spectrum of the doped semiconductors and $I_{0}$ is the average intensity of the undoped semiconductor. Compared with the undoped semiconductor $\left( v_{1}=0.37 \right) $, the average intensities of high-order harmonics of the doped semiconductors are all improved. It is also shown that the curves in Fig.5(a) and (b) have similar variation tendency. They both have maximums around $v_{1}=0.22$ and minimums around $v_{1}=0.37$. Figure 5(c) shows the band gaps e1 between VB and CB1 at different values of $v_{1}$. Figure 5(d) shows the band gaps e2 between CB2 and CB3 at different values of $v_{1}$. The minimums of e1 and e2 around $v_{1}=0.22$ in Fig.5(c) and Fig.5(d) correspond to the maximums of populations and intensities in Fig.5(a) and Fig.5(b), while the maximums of e1 and e2 around $v_{1}=0.37$ correspond to the minimums of populations and intensities.

Figure 5 shows that with the decreasing band gaps e1 and e2, the possibilities of electrons tunneling to higher energy bands become larger. Then the electron populations in the energy bands CB2 to CB3 and the intensity of the second plateau of the doped semiconductor high-order harmonics will be increased. The populations in CB2 to CB3 are primarily controlled by the band gap e1. Meanwhile, they can also be modulated by the band gap e2. Therefore one can control the species of dopants to control the semiconductor HHG, especially the second plateau.

We also investigate the HHG of the doped semiconductors under different doping rates at $v_{1}=0.52$ a.u. They are compared with that of the undoped semiconductor in Fig.6. It is shown in Figs.6(a) and (b) that the intensities of the high-order harmonics of the doped semiconductors between the 10th and the 32nd orders are about 0.5 times as those of the undoped semiconductor. While the intensity of the second plateau increases rapidly as the doping rate is risen from 0 to 0.1. In Figs.6(c) and (d), as the doping rate is larger than 0.1, the intensities of the high-order harmonics of the doped semiconductors between the 40th and the 70th orders are about 1 to 2 orders of magnitude higher than those of the undoped semiconductor.

\section{Conclusion}

We simulate the HHG in the doped and undoped semiconductors based on a 1D single-electron model in periodic potentials. The results indicate that the HHG in the semiconductors can be effectively controlled by doping. Both TDPI and energy bands picture are used to analyse the mechanism. Doping changes the energy band structure of the semiconductors and makes the Brillouin zone and band gaps e1 and e2 narrower than before. The small Brillouin zone and narrow band gaps e1 and e2 strengthen the electron populations in CB2 to CB3, and improve the intensity of the second plateau of the high-order harmonics. Our work indicates that one can control the HHG of the semiconductor by controlling the species of dopants and the doping rate.

\section{Acknowledgement}

This work was supported by the National Natural Science Foundation of China under Grants No. 11234004, No. 11404123, No. 11574101, No. 11422435, and No. 11627809.

\end{document}